# Space-time inverse-scattering of translation-based motion

Jeongsoo Kim[1], Shwetadwip Chowdhury[1,*]

**Authors information:**
[1] Department of Electrical and Computer Engineering, University of Texas at Austin, 2501 Speedway, Austin, TX 78712, USA
**\*Corresponding Author:** shwetadwip.chowdhury@utexas.edu

## Abstract

In optical diffraction tomography (ODT), a sample's 3D refractive-index (RI) is often reconstructed after illuminating it from multiple angles, with the assumption that the sample remains static throughout data collection. When the sample undergoes dynamic motion during this data-collection process, significant artifacts and distortions compromise the fidelity of the reconstructed images. In this study, we develop a space-time inverse-scattering technique for ODT that compensates for the translational motion of multiple-scattering samples during data collection. Our approach involves formulating a joint optimization problem to simultaneously estimate a scattering sample's translational position at each measurement and its motion-corrected 3D RI distribution. Experimental results demonstrate the technique's effectiveness, yielding reconstructions with reduced artifacts, enhanced spatial resolution, and improved quantitative accuracy for samples undergoing continuous translational motion during imaging.

# I. Introduction

Optical diffraction tomography (ODT) has emerged as a powerful class of optical imaging techniques for reconstructing three-dimensional refractive-index (RI) distributions within various samples[1-4]. This technique offers non-invasive, label-free, and quantitative contrast imaging, which is particularly valuable for studying the internal 3D structure of specimens without perturbing them. Consequently, ODT has achieved significant success in fields such as cell [5-7] and plant biology [8,9], neuroscience [10-12], pathology [13-15], and more.

Despite its significant advantages, traditional ODT faces two notable challenges: (1) *Its application has been predominantly limited to samples that are <u>weakly scattering</u>*. This limitation stems from the fact that traditional ODT reconstruction algorithms utilize weak-scattering models based on first-Born or Rytov approximations, which dramatically simplify the general scattering process [16-18]; and (2) Reconstructing a sample's 3D RI requires illuminating it from hundreds of different angles, a procedure that can take several seconds [19]. *During this time, any movement of the sample can introduce artifacts into the reconstructed RI distribution, compromising the accuracy and fidelity of the resulting images.*

Recently, advancements in ODT reconstruction algorithms have introduced more sophisticated scattering models (e.g., multi-slice beam propagation (MSBP) [20-23], split-step non-paraxial [24,25], wave propagation [26,27], Lippmann-Schwinger [28,29], multi-series born [30,31], etc.), which have higher accuracy when modelling scattering effects beyond the traditional weak-scattering approximation. By accounting for higher-order scattering, these models may broaden the scope of ODT to include imaging more complex, multicellular specimens that are not natively weak-scattering. Recent studies have shown preliminary but impressive 3D imaging results with small organisms, embryos, organoids, and thick-tissue specimens [22,30,32-34]. However, similar to their predecessors, image reconstruction frameworks based on these new scattering models still require multiple measurements to be collected. This challenge is particularly significant because these frameworks use nonlinear and nonconvex iterative solvers to "invert" the scattering models by searching a large nonconvex solution space for the optimal solution to the sample's 3D RI. Previous studies [35,36] have demonstrated that the likelihood of these inverse-scattering solvers to find this optimal solution increases with the number of measurements. Unfortunately, this implies that the more complex a sample's scattering is, the larger and more nonconvex the solution space becomes, which in turn requires a greater number of measurements (i.e., longer total acquisition time) to achieve robust 3D RI reconstruction. Currently, the scattering models introduced above assume the sample remains stationary during these measurements. Nonlinear and nonconvex inverse-solvers are particularly sensitive to this assumption, and even slight sample movements during the measurement sequence can result in severe motion artifacts or instabilities in the convergence process. Developing image reconstruction frameworks that can accommodate sample movement within a measurement sequence would dramatically broaden the scope of ODT imaging methods for dynamic, multiple-scattering samples.

Accommodating this type of inter-frame motion is not inherently a new idea in optical microscopy, especially for phase-imaging techniques that require multiple measurements to synthesize a single image [37-48]. For example, a popular non-interferometric phase-imaging method is differential phase contrast (DPC) microscopy, which reconstructs a sample's phase map using four intensity images [49,50]. Each image is captured with the sample illuminated in the far-field by a half-circle illumination pattern oriented at a different rotation angle, programmed onto an LED array. Any inter-frame sample motion occurring between these four measurements is challenging to isolate and correct, as intensity fluctuations in the measurements are influenced by both the sample features and changes in the illumination direction. To address this issue, recent work by Kellman et al. [42] leveraged the multi-colored programmable illumination capabilities of commercial LED arrays. They collected standard DPC measurements while also using a separate color channel on the LED array to simultaneously provide constant "navigator" illumination. The sample motion estimated from the navigator measurements was then corrected in the DPC measurements before image reconstruction. Later work by Gao et al. [41] demonstrated a more generalized phase-imaging framework that was able to correct for inter-frame sample motion without requiring separate navigator illumination. To accomplish this, they leveraged the fact that samples undergoing smooth and continuous motion exhibit inherent frame-to-frame similarity in their sample profile. This type of time-based sparsity was used to improve the conditioning of the computational phase-retrieval problem. By integrating a spatiotemporal total variation (TV) regularizer as a computational sparsity-promoting prior into the phase-retrieval algorithm, Gao et al. was able to promote smoothly varying temporal fluctuations in the reconstructed phase video, effectively correcting for inter-frame motion artifacts. A similar approach was presented by Sun et al., [51] who also utilized space-time priors to correct for inter-frame motion, and achieved dynamic phase imaging in the context of multiplexed Fourier ptychography. Specifically, Sun et al. formulated a space-time prior that applied a warping function to smoothly distort a time snapshot of the object based on the motion-field at that moment. By incorporating with the brightness constancy assumption from optical flow [52], this prior stabilized a joint-variable optimization process that reconstructed space-time functions for both the sample and its motion-fields, ensuring consistency with the inter-frame motion observed in the raw measurements.

This concept was recently extended towards 3D ODT by Qi et al. [53], who combined standard in-line holography with a 3D space-time optimization-based framework utilizing physics priors, to achieve high-speed 3D particle imaging velocimetry. Specifically, Qi et al. incorporated time-sparsity and flow incompressibility as physical priors to recover volumetric distributions of moving particles. Notably, in this demonstration, the sample was sparse, and the reconstruction framework exhibited more errors as the sample's sparsity decreased. Later work by Luo et al. [54] demonstrated a ODT space-time framework for non-sparse samples that used a multilayer perceptron, combined with disorder-invariant hashing and spatiotemporal regularization, to correct for 3D inter-frame motion from multiplexed measurements. This framework enabled volumetric reconstruction of dynamic, weak-scattering samples, and was demonstrated in simulation by imaging 3D phantom cells undergoing non-rigid deformation, as well as in experiment by volumetrically imaging oral epithelial cells undergoing controlled translational motion. Interestingly, neural networks have also been used to develop single-shot ODT methods, where fully 3D refractive index (RI) is reconstructed from a single measurement. For example, Ge et al. [55] demonstrated single-shot cell tomography for applications in high-speed cytometry. By illuminating a sample at multiple simultaneous directions, Ge et al. was able to multiplex various components of the sample's angular scattering information into a single interferometric measurement. A neural network, trained on a paired dataset of multiplexed measurements and corresponding 3D RI maps, was then used to computationally reconstruct the 3D RI from the measurement. This approach inherently eliminates frame-to-frame artifacts – however, imaging performance depends on how well the training dataset captures features in the sample to be imaged.

Other works in optical imaging, even beyond phase-imaging and ODT, have utilized spatiotemporal reconstruction frameworks to reconstruct continuous motion from multi-shot measurements [44,48,56]. For example, super-resolution microscopy techniques based on single molecule localization (SML) require the accumulation of thousands of localizations from multiple frames to construct a single super-resolved image. This process is time-consuming and easily degraded by inter-frame motion artifacts, significantly limiting the utility of these techniques in visualizing dynamic events [57,58]. To address these challenges, Saguy et al. [44] introduced a bi-directional long short-term memory (LSTM) network to achieve spatiotemporal super-resolution reconstruction. The LSTM network was designed to learn long-term structural and temporal correlations characteristic of individual samples by leveraging prior information about the sample type. As it processes each frame, the LSTM accumulates information from previous frames in its hidden state, progressively refining the reconstruction and capturing the sample's long-term characteristics. With super-resolved localization maps of a dynamic sample provided as input, the network performed spatiotemporal interpolation between frames, generating a motion-corrected sequence of super-resolved images. Structured illumination microscopy (SIM) is another type of super-resolution imaging that also requires multiple measurements, typically captured with sinusoidal illuminations with varying rotations, phase-shifts, and frequencies [59-63]. Though SIM requires far fewer measurements than SML and can thus capture faster dynamics [61,64], inter-frame motion can still degrade imaging performance [65,66]. To address this challenge, Cao et al. [56] proposed an inter-frame motion correction method that leveraged two coordinated-based neural networks to 1) estimate sample motion and 2) generate a motion-adjusted sample image. Specifically, the motion network produces a motion kernel at a specific timepoint, estimating the motion displacement for each pixel in the sample image. This adjusted motion information is then used by the image-generation network to estimate the sample function based on the motion-adjusted spatial coordinates. Based on the imaging parameters, this sample estimate is then used to estimate a raw measurement. In the learning process of these two networks, the network weights are optimized so that the difference between measurement estimates and raw measurements are minimized. At this point, both the motion fields and the motion-corrected sequence of the dynamic object will have been recovered. Cao et al. [56] demonstrated that this general framework can be applied to other imaging modalities as well, including phase-imaging.

  Though impressive, the methods described above have been demonstrated mainly for 2D samples or 3D samples that are either naturally sparse or weakly scattering. *To the best of our knowledge, there has not yet been any demonstration of inter-frame motion correction while also volumetrically solving the inverse-scattering problem for optically scattering, non-sparse samples.* This is a critical application gap since many multicellular samples (e.g., organ systems, small organisms, etc.) are optically scattering and highly dynamic at speeds beyond the frame-rates of standard ODT systems. These types of samples cannot currently be visualized with typical ODT imaging frameworks. *In this work, we develop a spatiotemporal scattering model for ODT that reconstructs the 3D RI of a scattering sample as it undergoes lateral translation motion between subsequent angular measurements.* Given the complexity of even just the spatial inverse-scattering problem, we have developed our spatiotemporal scattering model to use a purely physics-based formulation that parameterizes translational movement via joint-variable space-time estimation. This approach minimizes the number of unknown variables to iteratively solve for while also eliminating the need for training. Additionally, we chose not to incorporate any temporal priors beyond basic total variation (TV) regularization. We demonstrate and validate this framework by reconstructing 3D RI of scattering microphantom samples undergoing calibrated lateral translation motion between angular scattering measurements captured with a non-interferometric ODT system. This straightforward computational method for accounting for frame-to-frame translational movement between ODT measurements can enable the reconstruction of 3D RI in dynamic scattering samples where the primary motion is purely translational, such as in microfluidics [67,68], perfusion [69,70], and flow cytometry [71,72].

## 2. Computational framework

## A. Frame-to-frame motion imaged with multi-angle illumination

To correct for inter-frame translational motion artifacts, it is essential to recover the sample's trajectory. While standard image registration methods [73,74] could theoretically recover a sample's lateral displacements from raw measurements, they typically assume constant illumination. Under constant lighting, the same features are highlighted across frames, meaning intensity fluctuation in the measurements are solely due to sample movement. This scenario is ideal for standard registration algorithms. However, in ODT imaging, where angular scattering information is captured using various illumination angles, intensity variations will depend on both sample movement and illumination angle. We illustrate this in Figure 1, where the second and third rows compare simulated amplitude measurements of a linearly translating 3D sample under constant and multi-angle illumination, respectively. In the second row, with constant illumination, calculating the sample's frame-to-frame trajectory using standard registration methods is straightforward. However, in the third row, where different sample features are highlighted at various illumination angles, standard registration algorithms will struggle to determine the trajectory accurately. This is because the intensity variations they rely on are not dependent on only the sample's displacement. To overcome this limitation, we develop a space-time inverse-scattering framework based on joint-optimization, which reconstructs a scattering sample's 3D RI as well as its translational motion trajectory. In this work, we assume that the translational motion is fully lateral (2D), with no axial defocusing or rotation. We describe the specific mathematical formulation of this framework in the next section.

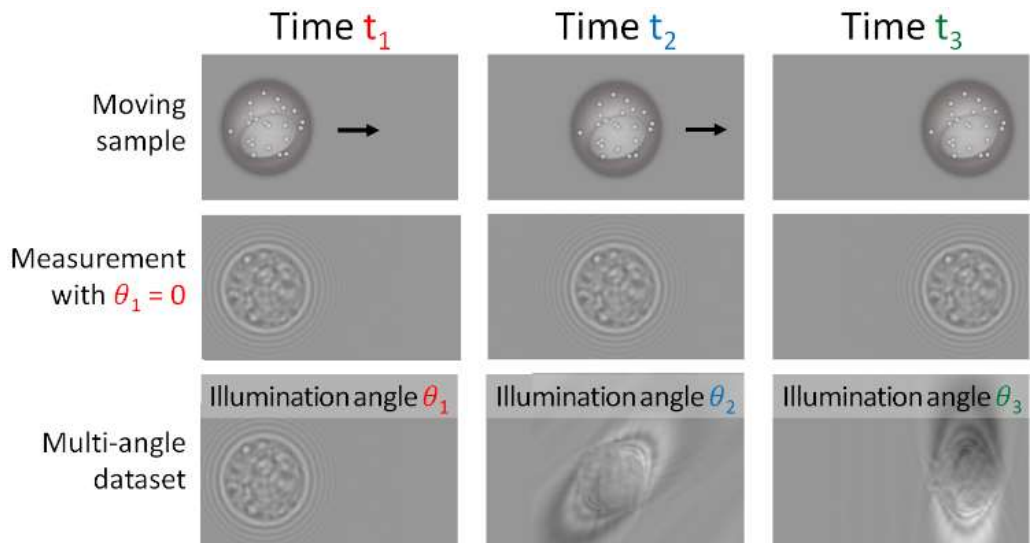

**Figure 1. (top row)** Three snapshots of a 3D scattering sample translating across a field-of-view, captured at time-points $t_1, t_2, t_3$. Simulated 2D raw measurements are shown with **(middle row)** constant illumination ($\theta_1 = 0°$), and **(bottom row)** varying illumination angles. Note that with changing illumination angle, different features of the sample are highlighted.

## B. Multi-slice beam-propagation forward model

We have recently designed a computational ODT framework based on the multi-slice beam propagation (MSBP) scattering model, which demonstrated promising results for unscrambling moderate amounts of optical scattering from a series of angular scattering measurements [22]. We extend this model to accommodate for translation motion occurring between measurement frames. To describe this, we first review the basics of our MSBP scattering model.

MSBP models light scattering through a 3D sample by dividing the sample into infinitesimally-thin layers, each of which individually acts as a phase mask that alters the phase of the propagating light wave. As shown in Figure 2, the sample's overall scattering characteristics are approximated by accumulating the diffraction effects that occur as light propagates layer-by-layer through the sample.

Consider a 3D sample being divided into 2D layers spaced $\Delta z$ distance apart. Mathematically, MSBP describes the evolution of the propagating light field by relating the light's electric-field at a given sample layer to the electric-field at the previous layer,

$$E_k(\boldsymbol{r}) = \exp\left(j \cdot \frac{2\pi}{\lambda} \cdot \Delta z \cdot [n_k(\boldsymbol{r}) - n_m]\right) \cdot H_{\Delta z}\{E_{k-1}(\boldsymbol{r})\} \qquad (1)$$

In Eq. (1), the vector $\boldsymbol{r}$ represents 2D $(x, y)$ lateral spatial coordinates, while $E_k(\boldsymbol{r})$ denotes the 2D electric-field exiting the $k$-th layer of the sample. The operator $H_{\Delta z}\{\cdot\}$ describes field propagation over a distance $\Delta z$, which we implement via angular propagation [75,76]. The term $\exp\left(j\frac{2\pi}{\lambda}\Delta z[n_k(\boldsymbol{r}) - n_m]\right)$ represents the transmittance function for the $k$-th sample layer, and depends on the difference between the 2D RI of the sample's $k$-th layer $n_k(\boldsymbol{r})$ and the homogenous media surrounding the sample $n_m$.

Consider the sample is decomposed into a total of $M$ layers. If the incident illumination field $E_0(\boldsymbol{r})$ and the sample's full 3D RI profile is known, then Eq. (1) can be used to propagate the incident field through the sample layer-by-layer to the sample's $M$-th final layer. The MSBP forward model, denoted by $\mathcal{G}\{.\}$, computes the resulting intensity measurement captured by a camera, can thus be written as,

$$\mathcal{G}\{n(\boldsymbol{r}_{3D})\} = |h(\boldsymbol{r}) \otimes H_{-\hat{z}}\{E_M(\boldsymbol{r})\}|^2 \qquad (2)$$

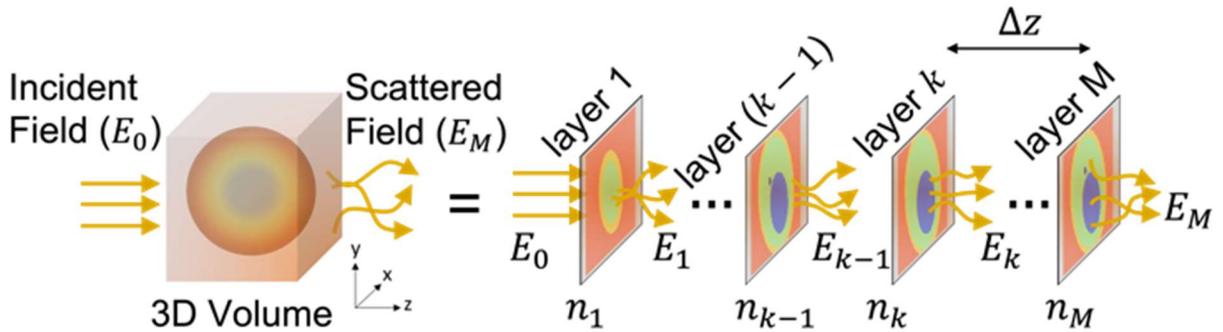

**Figure 2.** Multi-slice beam propagation (MSBP) models a 3D sample by breaking it down into a series of 2D layers. 3D scattering is modeled as a sequence of phase-modulation interactions as the incident light field ($E_0$) interacts with and passes through each layer till the last ($E_1, E_2, \ldots E_M$).

Here, $H_{-\hat{z}}\{E_M(\mathbf{r})\}$ indicates that the field emerging from the final $M$-th layer of the sample is back-propagated by a distance $\hat{z}$ back to the plane within the sample volume that is conjugate focused to the camera plane. This back-propagated field is then convolved with the system's 2D point-spread-function (PSF), $h(\mathbf{r})$, before being amplitude-squared to account for the fact that our optical system (described below) captures non-interferometric intensity measurements. We note that Eq. (2) expresses our MSBP forward model as a function of the sample's 3D RI $n(\mathbf{r}_{3D})$, where $\mathbf{r}_{3D}$ represents 3D spatial coordinates $(x, y, k)$ that discretizes the sample's axial dimension based on the layer-to-layer decomposition illustrated in Figure 2, i.e., $n(\mathbf{r}_{3D}) = n_k(\mathbf{r})$ for $k = 1,2,\ldots M$.

## C. Space-time multi-slice beam-propagation for translation-based motion

As discussed previously, we collect multiple measurement of the sample under varying illumination patterns. We denote the incident fields corresponding to each pattern as $E_0^{(\ell)}(\mathbf{r})$, where $\ell = 1,2,\ldots,L$ represents each measurement within the total dataset, captured at equally-spaced time intervals. If the sample moves between subsequent measurements, the parameter $\ell$ can be integrated into the RI term of Eq. (1) to indicate that the sample's RI distribution varies with each measurement, i.e., $n_k^{(\ell)}(\mathbf{r})$. While one could attempt to solve for $n_k^{(\ell)}(\mathbf{r})$ as an independent distribution for each measurement, this strategy would significantly increase the number of unknown degrees-of-freedom in an inverse-problem already ill-posed due to the complexity of scattering. Although spatiotemporal priors [77] can be used to constrain the solution-space for $n_k^{(\ell)}(\mathbf{r})$, none have yet been effectively demonstrated for space-time inverse-scattering of dynamic non-sparse scattering samples. To avoid these challenges and leverage the fact that we are *only* considering 2D $(x, y)$ translational sample motion, we instead parameterize frame-to-frame translation by introducing a displacement vector $\delta \mathbf{r}^{(\ell)}$,

$$n_k^{(\ell)}(\mathbf{r}) = n_k(\mathbf{r} - \delta \mathbf{r}^{(\ell)}) \quad (3)$$

Here, $\delta \mathbf{r}^{(\ell)}$ represents a 2-element lateral displacement vector that describes how the sample's native RI distribution $n_k(\mathbf{r})$ was shifted by the time the $\ell$-th measurement was captured. As a result, the beam propagation and phase multiplication step in Eq. (1), along with the formulation of the MSBP forward model in Eq. (2), can be updated to incorporate the measurement parameter $\ell$ and the parameterized description of $n_k^{(\ell)}(\mathbf{r})$ from Eq. (3),

$$E_k^{(\ell)}(\mathbf{r}) = \exp\left(j \cdot \frac{2\pi}{\lambda} \Delta z \cdot [n_k(\mathbf{r} - \delta \mathbf{r}^{(\ell)}) - n_m]\right) \cdot H_{\Delta z}\{E_{k-1}^{(\ell)}(\mathbf{r})\} \quad (4)$$

$$\mathcal{G}\{n(\mathbf{r}_{3D}), \delta \mathbf{r}^{(\ell)}\} = \left|h(\mathbf{r}) \otimes H_{-\hat{z}}\{E_M^{(\ell)}(\mathbf{r})\}\right|^2 \quad (5)$$

In Eq. (4), $E_k^{(\ell)}(\mathbf{r})$ and $E_{k-1}^{(\ell)}(\mathbf{r})$ represent the fields exiting the $k$-th and $(k-1)$-th sample layers, respectively, in response to the $\ell$-th incident field $E_0^{\ell}(\mathbf{r})$. It follows that the updated forward model $\mathcal{G}\{.\}$ shown in Eq. (5) now accounts for the measurement parameter $\ell$, and takes as input variables the sample's native 3D RI distribution $n(\boldsymbol{r}_{3D})$ *and* the $\ell$-th displacement vector $\delta \boldsymbol{r}^{(\ell)}$, and outputs a prediction of the $\ell$-th intensity measurement of a translating sample. Recall that $n(\boldsymbol{r}_{3D})$ is a 3D *spatial* distribution, while $\delta \boldsymbol{r}^{(\ell)}$ denotes the sample's 2D translational displacement at a specific *time* point. *Thus, Eq. (5) presents $\mathcal{G}\{.\}$ as the <u>space-time MSBP forward model for translation-based motion</u>.* Notably, if $\delta \boldsymbol{r}^{(\ell)} = 0$ for all $\ell$, then Eqs. (4) and (5) revert to the purely spatial MSBP model described by Eqs. (1) and (2).

## D. Inverse problem formulation

In practice, we collect multiple-scattering measurements of the sample being illuminated by angled plane-waves. The incident fields associated with these illumination plane-waves can be expressed as $E_0^{(\ell)}(\boldsymbol{r}) = \exp(j\,\boldsymbol{k}_0^{(\ell)} \cdot \boldsymbol{r})$, where $\boldsymbol{k}_0^{(\ell)}$ describes the 2D $(k_x, k_y)$ lateral wave-vector associated with the $\ell$-th plane-wave illumination, which in turn determines the wave's angle of incidence on the sample. We denote the corresponding $\ell$-th intensity measurement as $I^{(\ell)}(\boldsymbol{r})$.

We now formulate a *joint-variable* reconstruction framework to simultaneously determine the sample's native RI distribution and its translational displacement at each illumination angle. By jointly solving for these variables, the reconstruction of the scattering sample's 3D RI will inherently account for the sample's frame-to-frame translational motion. We formulate this reconstruction framework as a joint-variable least-squares minimization that searches for an optimal solution for the sample's 3D RI and translation trajectory that minimizes the difference between the measurement amplitudes (square-root of intensity) and the amplitude predictions made by the space-time MSBP forward model,

$$( \hat{n}(\boldsymbol{r}_{3D}), \Delta \hat{\boldsymbol{r}} )$$
$$= \underset{n(\boldsymbol{r}_{3D}), \Delta \boldsymbol{r}}{\arg\min} \sum_{\ell=1}^{L} \left\| \sqrt{I^{(\ell)}(\boldsymbol{r})} - \sqrt{\mathcal{G}\{n(\boldsymbol{r}_{3D}), \delta \boldsymbol{r}^{(\ell)}\}} \right\|_{L2}^{2} \qquad (6)$$
$$+ \alpha \cdot \text{TV}_{x,y,z}\{n(\boldsymbol{r}_{3D})\} + \beta \cdot \text{TV}_t\{\Delta \boldsymbol{r}\}$$

Here, $\|...\|_{L2}$ denotes the L2-norm operator [78] and $\Delta \boldsymbol{r}$ aggregates all the individual $(x, y)$ displacement vectors into one matrix variable, i.e., $\Delta \boldsymbol{r} = (\delta \boldsymbol{r}^{(1)}; \delta \boldsymbol{r}^{(2)}; ... \delta \boldsymbol{r}^{(\ell)}; ...; \delta \boldsymbol{r}^{(L)})$. In this notation, the $\ell$-th row of $\Delta \boldsymbol{r}$ correspond to the $x$ and $y$ coordinates of $\delta \boldsymbol{r}^{(\ell)}$. Thus, the rows of $\Delta \boldsymbol{r}$ effectively represent sample displacement over the time-axis.

The first term in Eq. (6) within the L2 operator assesses how well the current estimates of $n(\boldsymbol{r}_{3D})$ and $\Delta \hat{\boldsymbol{r}}$ allow the forward model to match the experimental scattering measurements. Standard descent-based algorithms can be employed to iteratively refine these estimates, optimizing the alignment between the model and the data. To guide this optimization and enhance stability, the second and third terms in Eq. (6) incorporate prior knowledge of spatiotemporal scattering dynamics. Specifically, we leverage the fact that most real-world dynamic samples exhibit smooth, spatially-varying profiles and continuous motion. Therefore, we apply the total-variation (TV) regularizer [79,80] as a spatiotemporal sparsity-promoting prior. Specifically, the second term $\text{TV}_{x,y,z}\{\hat{n}(\boldsymbol{r}_{3D})\}$ represents TV regularization across the 3D spatial distribution of $n(\boldsymbol{r}_{3D})$ while the third term $\text{TV}_t\{\Delta \boldsymbol{r}\}$ regularizes $\Delta \boldsymbol{r}$ across the time dimension. Together, Eq. (6) effectively performs 4D space-time optimization, and simultaneously reconstructs the scattering sample's volumetric RI while also correcting for the sample's frame-to-frame translation-based motion. $\alpha$ and $\beta$ denote the regularization strengths of the spatial and temporal TV regularizers, respectively.

To start the descent-based optimization, $n(\mathbf{r}_{3D})$ and $\Delta\mathbf{r}$ have to first be initialized. Similar to our previous work [22], we initialized $n(\mathbf{r}_{3D})$ to be uniformly equal to $n_m$. However, to initialize $\Delta\mathbf{r}$, three separate methods were employed: 1) standard image registration [73] between raw measurements to estimate frame-to-frame sample displacement, despite variations in illumination angle across measurements; 2) identification of centroids for each measurement following background-subtraction and binarization [81], which were then compared to estimate motion between frames. We observed that centroids were more robust to changes in illumination angle (see Supplementary document); and 3) external feedback from the motorized linear stage controlling sample motion (see Fig 3 below), which directly estimated sample's frame-to-frame displacement. Due to the nonconvexity of the space-time inverse-scattering problem, we observed from simulations (see Supplementary document) that the quality of space-time reconstruction was dependent on how we chose to initialize $\Delta\mathbf{r}$. We also demonstrate this with real-world experiments, as described below.

# 3. Experimental results
## A. Optical setup
The schematic of our non-interferometric ODT imaging setup is shown in Figure 3. Light from a red laser (Thorlabs, CPS635R, center wavelength: $\lambda$ = 635 nm) is coupled into the setup through a single-mode fiber. The output light from the fiber is collimated using an objective lens (L1, Newport, M-10X, NA: 0.25) and angularly steered by a dual-axis galvo-mirror (GM, Thorlabs, GVS012). The system's first 4f system, composed of an achromat lens (L2, Thorlabs, ACT508-200-A) and an imaging objective (OBJ, Nikon, CFI Plan Apochromat Lambda D 60X Oil, NA: 1.42), conjugates the plane of the galvo-mirror into the sample volume. Changing the angle of the GM steers the illumination angle incident to the sample, enabling both on-axis and off-axis illumination of the sample. The light diffracted from the sample is collected by a second 4f system identical to the first. Unlike in our previous work [22], where a physical iris was placed in the sample's Fourier plane to limit the system's effective numerical aperture (NA) and reduce aberrations, we found empirically that this step was unnecessary with our choice of OBJ. Consequently, we use the second 4f system to directly image light from the sample onto the CMOS camera (ToupTek, IUA20000KMA), which non-interferometrically measures intensity. Since the system operates at the full NA = 1.42 allowed by OBJ, our theoretical lateral and axial resolutions are given by $\lambda/2NA = 223.5$ nm and $2\lambda/NA^2 = 629.8$ nm [82], respectively. Lastly, the sample is positioned on a 2-axis motorized translation stage, which enables continuous 2D lateral movement.

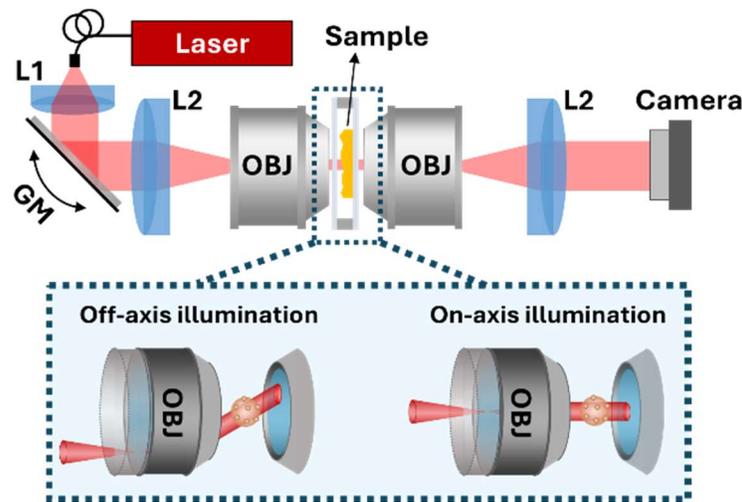

**Figure 3.** Schematic of experimental optical setup. (L: Lens, GM: Galvo mirror, OBJ: Objective lens). Sample is mounted on a motorized translation stage (not shown).

## B. Space-time inverse-scattering on microsphere sample undergoing translation motion

To demonstrate our space-time inverse-scattering framework, we first reconstruct 3D RI in a sample consisting of 10 μm diameter polystyrene microspheres immersed in index-matched oil (see Figure 4 above). The microspheres were immersed in index-matching oil (Cargille Standard Series Liquids) with RI of $n_m = 1.5682$ at the 635 nm center wavelength of our light source. Given that polystyrene has an RI of $n_{PS} = 1.5874$ at the same wavelength, the "ground-truth" microsphere-to-media RI difference for the polystyrene microspheres was $\Delta n_g = n_{PS} - n_m = 0.0192$.

The sample was illuminated sequentially from 200 different angles arranged in a spiral trajectory. For each angle, the intensity of the resulting scattering measurements was acquired. Data collection was conducted under two scenarios: 1) the sample remained stationary throughout the acquisition process, and 2) the sample underwent continuous translational motion between measurements, driven by a motorized translation stage following a square spiral pattern. In the stationary scenario, we applied the traditional MSBP framework as a space-only inverse scattering approach to reconstruct the sample's 3D RI. We consider this as the baseline standard, representing the best-case scenario for reconstructing the sample's RI. In the scenario when the sample is moving, we first performed space-only inverse scattering by solving Eq. (6) as a single-variable optimization problem for $n(\boldsymbol{r}_{3D})$, where the displacement vectors in $\Delta \boldsymbol{r}$ (e.g., motion trajectory) were fixed based on values computed through frame-to-frame registration, external feedback from the motorized stage, or centroid localization. We then conducted fully space-time inverse scattering in this scenario by solving Eq. (6) as a joint-variable optimization problem. The previously computed displacement vectors in $\Delta \boldsymbol{r}$, as described earlier, were used to initialize the process for iteratively optimizing $\Delta \boldsymbol{r}$. In all cases, the distribution for the sample's 3D RI was initialized to be equal to the surrounding media.

In Figures 4(a-e) below, we show 3D RI reconstruction results for the various imaging and reconstruction conditions discussed previously. We also show plots comparing the RI-difference values ($\Delta n(\boldsymbol{r}_{3D}) = \hat{n}(\boldsymbol{r}_{3D}) - n_m$), indicated by colored solid lines, with the ground-truth RI-difference values within the microspheres predicted by $\Delta n_g$ (dashed black line). In Figure 4(a), where space-only inverse-scattering of a static sample was conducted, we see that the sample's circular microsphere structure is accurately reconstructed without noticeable artifacts, and the reconstructed RI-differences match well with ground-truth expectations (less than 1% error). However, based on Figures 4(b-d), space-only reconstruction of a dataset containing inter-frame sample-motion, even when motion is estimated a-priori, results in noticeable blur artifacts around the microspheres and incorrect RI-difference values. These artifacts are especially pronounced when sample displacement across time is estimated via registration (Fig. 4(b)) and centroid localization (Fig. 4(d)). However, even in the case when sample motion is determined by the direct feedback from the motorized stage, blurring artifacts are still observed. We point out these artifacts with yellow arrows in Figure 4(c). In contrast, Figure 4(e) shows that space-time inverse-scattering achieves highly accurate 3D RI reconstruction, free of blurring artifacts and with less than 1% RI error – virtually indistinguishable from the static sample reconstruction shown in Figure 4(a). Results shown in Figure 4(e) specifically used centroid localization to initialize the sample motion – however, we empirically observed virtually identical results regardless of which motion-initialization method was used (see Supplementary document).

Figure 4(f) compares the sample's motion trajectories obtained from frame-to-frame registration, motorized stage feedback, and centroid localization. These trajectories, obtained without joint space-time optimization, are also contrasted with the motion trajectory iteratively refined during joint space-time reconstruction, corresponding to the 3D RI reconstruction depicted in Figure 4(e). While the motorized stage feedback trajectory follows the programmed square spiral pattern, trajectories from frame-to-frame registration and centroid localization deviate substantially. Interestingly, the trajectory derived through joint space-time reconstruction also slightly deviates from the programmed stage trajectory. However, we recall that this slightly deviant trajectory yielded the best 3D RI reconstruction. This underscores the challenges of precisely controlling motion trajectories with a mechanical stage and highlights the need for computational optimization to enhance accuracy. These observed results remained consistent regardless of whether time-based regularization was applied to enforce trajectory smoothness during joint-variable optimization.

These findings demonstrate the feasibility of the proposed space-time inverse-scattering approach to accurately reconstruct 3D RI and motion trajectories for translational motion, at least in weakly scattering samples such as index-matched microspheres. We observed similar success with weakly scattering biological samples, such as cheek cells (see Supplementary document). We now demonstrate this method in more complex scattering samples.

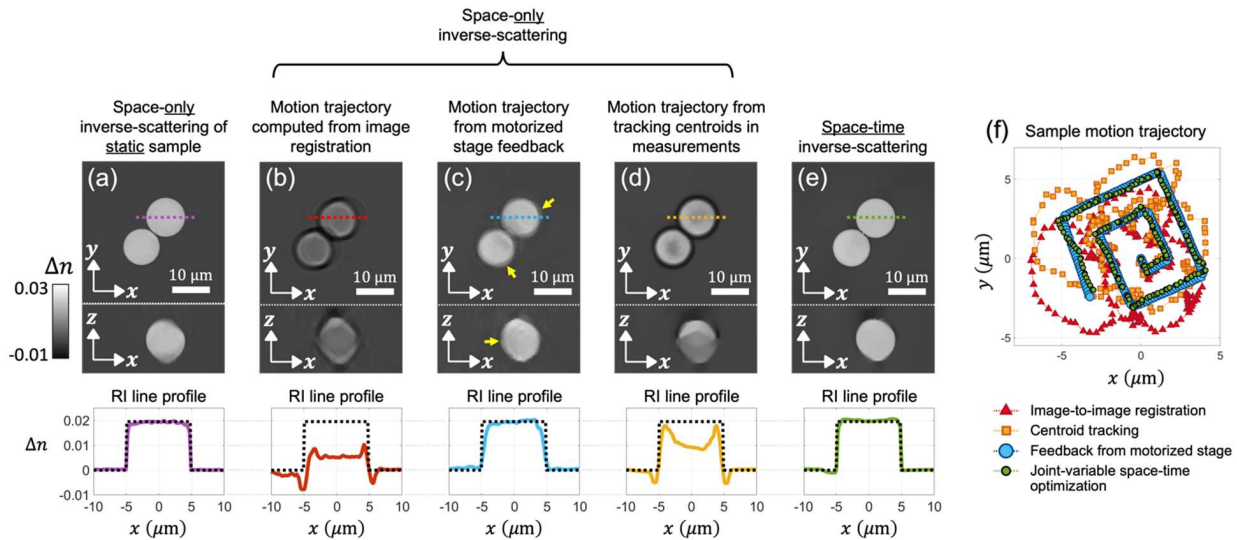

**Figure 4**. Comparison of space-only and space-time reconstruction results using static and moving microspheres. **(a–e)** Reconstructed 3D RI-difference images are shown in the lateral ($x - y$) and axial ($x - z$) planes. Furthermore, RI profiles across the dashed colored line in the x-y images compare reconstructed RI values to the ground truth RI difference (dotted black lines). **(a)** Space-only reconstruction of the static microsphere sample. **(b–d)** Space-only reconstructions of the moving sample after computing motion trajectories from image registration, motorized stage feedback, and centroid tracking, respectively. **(e)** Joint-variable space-time reconstruction of the moving sample, using centroid tracking to initialize the sample's motion trajectory before iterative update. **(f)** Motion trajectories of the sample are shown as estimated by different methods.

## C. Scattering phantom

To demonstrate whether our space-time inverse-scattering method is applicable to multiple-scattering samples, we conducted experiments on a scattering microphantom sample composed of a cell-mimicking target encased within a $40 \times 40 \times 40$ μm$^3$ scattering cube [83,84]. Scattering effects in this cube are induced by irregularly-spaced horizontal and vertical rods, each having 400 nm lateral width. We verified the presence of multiple-scattering effects in this sample by performing conventional interferometric ODT using the Rytov weak-scattering assumption for 3D reconstruction. As shown in the Supplementary document, the resulting reconstruction exhibits substantial artifacts that significantly compromise image quality.

     As in the previous experiment, angular scattering measurements were collected while the sample underwent translational motion along a square spiral trajectory. This time, a total of 300 illumination angles were used, and the sample's motion trajectory was again initialized based on either frame-to-frame registration, external feedback from the motorized stage, and centroid localization. 3D RI reconstruction results are shown in Figure 5 below.

     As the baseline standard, Figure 5(a) shows the traditional MSBP reconstruction for the 3D RI when the sample is kept static during measurements. Lateral and axial cross-sectional planes, as well as the tomographic view of the 3D RI, reveals that both the scattering cube and the cell-mimicking target inside are successfully reconstructed. Individual rods within the scattering cube as well as subcellular features within the target can be clearly visualized.

     Figures 5(b,c) present the inverse-scattering results for the sample undergoing translational square-spiral motion, with trajectory initialization estimates obtained from frame-to-frame registration and centroid localization, respectively. Specifically, Figures 5(b1, c1) show the sample's 3D RI reconstructed using single-variable space-only reconstruction. Figures 5(b2, c2) and (b3, c3) present the results of joint-variable space-time reconstruction without ($\beta = 0$) and with ($\beta = 0.01$) time-based total-variation (TV) regularization, respectively. Lastly, Figures 5(b4, c4) both compare the motion trajectories directly estimated through either frame-to-frame registration or centroid localization (red triangle), with those iteratively refined via space-time optimization without (orange square) and with (blue circle) time-based TV regularization.

     Figures 5(b1, c1) show that space-only inverse scattering failed to recover any sample-specific features, regardless of which trajectory estimation method was used. This suggests that the motion trajectories directly estimated through both frame-to-frame registration and centroid localization are inaccurate, as expected. Notably, when the motion trajectory is initialized using frame-to-frame registration, space-time inverse scattering (Figures 5(b2, b3)) also fails to recover any discernible sample features. However, this is not the case when centroid localization is used for trajectory initialization. In this scenario, Figures 5(c2, c3) demonstrate that bulk features of the microphantom are discernible. This improvement occurs because the motion trajectories estimated via centroid localization (Figure 5(c4)) are more accurate than those derived from frame-to-frame registration (Figure 5(b4)). This provides a *sufficiently* accurate initial guess for space-time optimization to then refine trajectory values effectively via joint-variable optimization. Indeed, a comparison of Figures 5(c4) and 5(b4) highlights that the trajectory estimation

based on centroid localization exhibits higher localization and structural coherence than the estimation obtained through frame-to-frame registration.

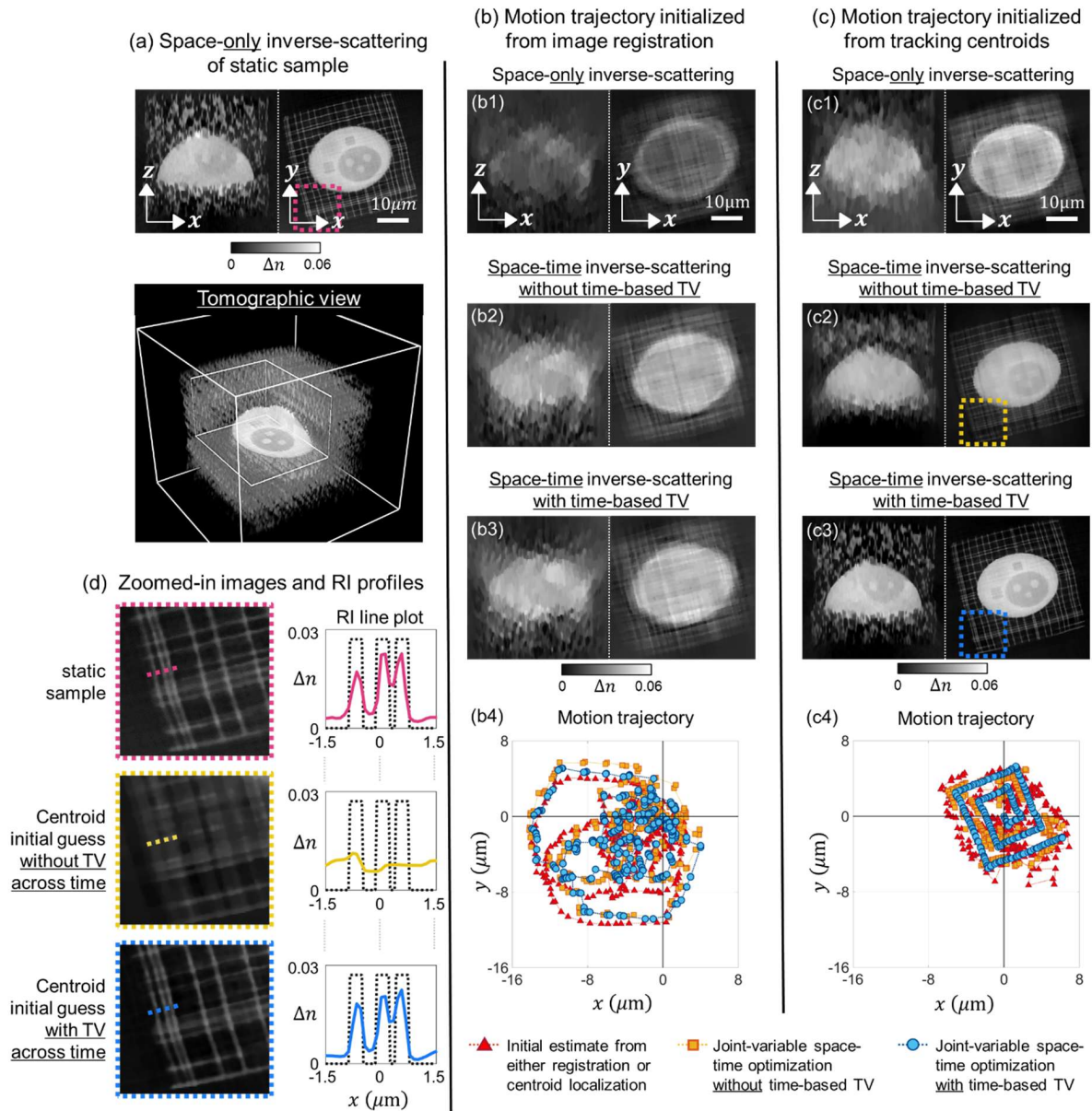

**Figure 5.** Reconstruction results of the scattering phantom under static and moving conditions. **(a)** Tomographic view and cross-sectional slices (x-z and x-y planes) of the 3D reconstruction of the static scattering phantom, clearly resolving internal features of the scattering cube and imaging target. **(b1-b3)** Reconstruction results for the moving sample using trajectory computed from frame-to-frame registration as the initial guess. Both space-only and space-time reconstructions fail to recover internal features, even with time-based regularization. **(b4)** The initial and reconstructed sample trajectories show substantial deviations from the designed square spiral pattern. **(c1-c3)** Reconstruction results using the centroid-based trajectory as the initial guess. While space-only reconstruction remains unsuccessful, space-time reconstruction with trajectory regularization successfully recovers high-resolution internal features. **(c4)** Comparison of initial and reconstructed motion trajectories for the centroid-based initialization method, showing that temporal regularization improves trajectory alignment to match the desired square spiral pattern. **(d)** Magnified images of the boxed regions in (a), (c2), and (c3) emphasizes the output spatial resolution of the various space-time reconstruction strategies. RI profiles taken across the dashed colored lines in the magnified images compare reconstructed RI-difference values with ground-truth (dotted black).

To further investigate reconstruction quality in Figures 5(c2, c3), we zoom in on a region containing scattering rods, which are the highest-resolution feature of this specific microphantom (indicated by the dashed rectangle in Figures 5(c2) and 5(c3)). Figure 5(d) compares close-up views of Figures 5(c2) and 5(c3) with the baseline standard (Figure 1(a)) and includes RI line plots. In the baseline 3D RI reconstruction of the static sample, the rods are clearly resolved, as shown in the zoom-in and corresponding RI line plot, where three distinct peaks represent individual rods. However, in the case of a translating sample, we see that though initializing the sample's motion via centroid localization allows bulk-feature visualization, not using time-based TV to promote motion continuity in the trajectory solution results in a recovered motion trajectory deviating significantly from the expected square spiral path (Figure 5(c4)). Consequently, imaging resolution is degraded and individual rods cannot be distinguished, as shown in the zoomed view in Figure 5(d). In contrast, when the time-based TV prior is incorporated, the optimization successfully refines the motion trajectory to recover the square spiral path accurately (Figure 5(c4)). This in turn enables high-resolution 3D RI reconstruction, virtually indistinguishable from the baseline standard (Figure 5(d)).

Notably, direct feedback from the motorized stage provides an even more accurate initial guess for the motion trajectory compared to centroid localization. With this improved initialization, space-time reconstruction successfully recovered both the square spiral motion trajectory of the sample and its high-resolution 3D RI, regardless of whether TV regularization was applied to the motion trajectory. Detailed results are provided in the supplementary document.

Building on our conclusions from the microsphere sample in Figure 4, the microphantom experiments reaffirm that space-time inverse scattering requires joint-variable reconstruction of both the sample's 3D RI and its motion trajectory. An important additional insight from these microphantom experiments is that the nonconvexity of this optimization problem increases with greater sample scattering. In such cases, having an accurate initial guess for the motion trajectory becomes crucial. In the case when a highly accurate initial guess can be obtained – such as from the high-precision feedback from a motorized stage – the joint-variable space-time optimization can achieve successful reconstruction without additional imaging priors. However, in scenarios where such precise prior knowledge is unavailable (e.g., when motion arises from inherent sample dynamics rather than controlled external movement), imaging priors become essential to constrain and guide the optimization process. In Figure 5(c) where centroid localization provided a less accurate initial guess for the trajectory than stage feedback, we achieved effective space-time reconstruction by applying time-based TV to the motion trajectory during optimization. However, when relying on even less accurate initialization methods, such as frame-to-frame registration as shown in Figure 5(b), even stronger imaging priors will likely be required to achieve reliable results.

## 4. Discussion and conclusion

In this study, we have introduced a novel space-time inverse-scattering method for ODT that accounts for translational motion that a scattering sample undergoes between frames during data acquisition. This method extends the traditional MSBP framework into a joint optimization process that simultaneously reconstructs the sample's 3D RI and determines its translational motion trajectory during measurement. Importantly, this method is computationally efficient and relies solely on physics-based optimization without the need for training, which enhances its robustness and practical applicability. We have demonstrated the effectiveness of our motion correction method by reconstructing both weak- and multiple-scattering samples, such as microspheres and microphantoms, respectively, as they undergo translational motion.

Inverse-scattering is inherently ill-posed problem, even for static samples. Introducing unknown sample movement between frames further exacerbates the problem. Thus, a well-informed initial guess for the sample's motion trajectory is critical to effectively initiate the joint-variable optimization process. Notably, the accuracy of this initial guess is heavily influenced by how scattering the sample is. For example, in our demonstration with weakly-scattering microspheres (Figure 4), both frame-to-frame registration and centroid localization provided sufficiently accurate trajectory estimates. However, in the case of more strongly scattering microphantoms (Figure 5), only centroid localization provided an estimate accurate enough to enable high-quality joint-variable reconstruction. Furthermore, this was only achieved if combined with time-based TV regularization, which enforced continuity in the reconstructed motion trajectory. As the nonconvexity of the space-time inverse-scattering problem further increases for even more scattering samples, more advanced strategies may be required that combine different motion-prediction strategies with more sophisticated image priors to achieve accurate trajectory initialization. We expect such strategies will need to be tailored to specific classes of sample scattering since their effectiveness is ultimately sample dependent.

Notably, this work focuses exclusively on translational sample motion. Accommodating for complex non-rigid motion between measurement frames presents another significant challenge for space-time inverse scattering. While recent studies (described earlier) have explored space-time frameworks to account for non-rigid motion, their real-world demonstrations have been limited to dynamic 2D samples [39-48, 51], weakly scattering 3D samples [54-56], or sparse 3D samples [53]. To our knowledge, robust space-time reconstruction for dynamic, non-sparse, and 3D scattering samples has yet to be achieved. We believe that the primary challenge in this lies in the extreme ill-posedness of the problem. Even in this study, handling inter-frame translational motion, which is efficiently parameterized by simple 2D displacement vectors, required both a well-informed initial guess for the sample's motion trajectory and TV-based imaging priors to constrain the joint-variable optimization. Non-rigid motion, with its inherent complexity, introduces a significantly larger number of unknown parameters into the inverse problem. Developing methods to constrain and solve this ill-posed and highly nonconvex problem will be an exciting and critical direction for future research.

**Funding**. This project has been made possible by grant number 2023-321173 and 2021-225666 from the Chan Zuckerberg Initiative DAF, an advised fund of Silicon Valley Community Foundation, as well as NIH grant R35GM155424. Additional support was also provided by the University of Texas at Austin (UT), Cockrell School of Engineering, and Chandra Family Department of Electrical and Computer Engineering.

**Acknowledgment**. We want to thank Michał Ziemczonok, Wojciech Krauze, Arkadiusz Kuś, and Małgorzata Kujawińska from Warsaw University of Technology for their assistance in 3D printing the scattering microphantom.

**Disclosures.** The authors declare that there are no conflicts of interest related to this article.

**Data availability.** Data underlying the results presented in this paper may be obtained from the authors upon reasonable request.

# References

1   Choi, W. *et al.* Tomographic phase microscopy. *Nature methods* **4**, 717-719 (2007).
2   Sung, Y. *et al.* Optical diffraction tomography for high resolution live cell imaging. *Optics express* **17**, 266-277 (2009).
3   Kim, T. *et al.* White-light diffraction tomography of unlabelled live cells. *Nature Photonics* **8**, 256-263 (2014).
4   Lee, K. *et al.* Quantitative phase imaging techniques for the study of cell pathophysiology: from principles to applications. *Sensors* **13**, 4170-4191 (2013).
5   Shin, S., Kim, D., Kim, K. & Park, Y. Super-resolution three-dimensional fluorescence and optical diffraction tomography of live cells using structured illumination generated by a digital micromirror device. *Scientific reports* **8**, 9183 (2018).
6   Sherman, D. J. *et al.* The fatty liver disease–causing protein PNPLA3-I148M alters lipid droplet–Golgi dynamics. *Proceedings of the National Academy of Sciences* **121**, e2318619121 (2024).
7   Saunders, N. *et al.* Dynamic label-free analysis of SARS-CoV-2 infection reveals virus-induced subcellular remodeling. *Nature Communications* **15**, 4996 (2024).
8   Park, C. *et al.* Three-dimensional refractive-index distributions of individual angiosperm pollen grains. *Current Optics and Photonics* **2**, 460-467 (2018).
9   Samalova, M. *et al.* Hormone-regulated expansins: expression, localization, and cell wall biomechanics in Arabidopsis root growth. *Plant Physiology* **194**, 209-228 (2024).
10  Cotte, Y. *et al.* Marker-free phase nanoscopy. *Nature Photonics* **7**, 113-117 (2013).
11  Yang, S. A., Yoon, J., Kim, K. & Park, Y. Measurements of morphological and biophysical alterations in individual neuron cells associated with early neurotoxic effects in Parkinson's disease. *Cytometry part A* **91**, 510-518 (2017).
12  Pengsart, W., Tongkrajang, N., Whangviboonkij, N., Sarasombath, P. T. & Kulkeaw, K. Balamuthia mandrillaris trophozoites ingest human neuronal cells via a trogocytosis-independent mechanism. *Parasites & Vectors* **15**, 232 (2022).
13  Hugonnet, H. *et al.* Multiscale label-free volumetric holographic histopathology of thick-tissue slides with subcellular resolution. *Advanced Photonics* **3**, 026004-026004 (2021).
14  Zadka, Ł. *et al.* Label-Free Quantitative Phase Imaging Reveals Spatial Heterogeneity of Extracellular Vesicles in Select Colon Disorders. *The American Journal of Pathology* **191**, 2147-2171 (2021).
15  Park, J. *et al.* Quantification of structural heterogeneity in H&E stained clear cell renal cell carcinoma using refractive index tomography. *Biomedical Optics Express* **14**, 1071-1081 (2023).
16  Wolf, E. Three-dimensional structure determination of semi-transparent objects from holographic data. *Optics communications* **1**, 153-156 (1969).
17  Devaney, A. Inverse-scattering theory within the Rytov approximation. *Optics letters* **6**, 374-376 (1981).
18  Born, M. & Wolf, E. *Principles of optics: electromagnetic theory of propagation, interference and diffraction of light*.  (Elsevier, 2013).

19 Balasubramani, V. *et al.* Holographic tomography: techniques and biomedical applications. *Applied Optics* **60**, B65-B80 (2021).
20 Tian, L. & Waller, L. 3D intensity and phase imaging from light field measurements in an LED array microscope. *optica* **2**, 104-111 (2015).
21 Kamilov, U. S. *et al.* Optical tomographic image reconstruction based on beam propagation and sparse regularization. *IEEE Transactions on Computational Imaging* **2**, 59-70 (2016).
22 Chowdhury, S. *et al.* High-resolution 3D refractive index microscopy of multiple-scattering samples from intensity images. *Optica* **6**, 1211-1219 (2019).
23 Chen, M., Ren, D., Liu, H.-Y., Chowdhury, S. & Waller, L. Multi-layer Born multiple-scattering model for 3D phase microscopy. *Optica* **7**, 394-403 (2020).
24 Lim, J., Ayoub, A. B., Antoine, E. E. & Psaltis, D. High-fidelity optical diffraction tomography of multiple scattering samples. *Light: Science & Applications* **8**, 82 (2019).
25 Zhu, J., Wang, H. & Tian, L. High-fidelity intensity diffraction tomography with a non-paraxial multiple-scattering model. *Optics Express* **30**, 32808-32821 (2022).
26 Ma, X., Xiao, W. & Pan, F. Optical tomographic reconstruction based on multi-slice wave propagation method. *Optics express* **25**, 22595-22607 (2017).
27 Suski, D., Winnik, J. & Kozacki, T. Fast multiple-scattering holographic tomography based on the wave propagation method. *Applied Optics* **59**, 1397-1403 (2020).
28 Liu, H.-Y. *et al.* SEAGLE: Sparsity-driven image reconstruction under multiple scattering. *IEEE Transactions on Computational Imaging* **4**, 73-86 (2017).
29 Soubies, E., Pham, T.-A. & Unser, M. Efficient inversion of multiple-scattering model for optical diffraction tomography. *Optics express* **25**, 21786-21800 (2017).
30 Lee, M., Hugonnet, H. & Park, Y. Inverse problem solver for multiple light scattering using modified Born series. *Optica* **9**, 177-182 (2022).
31 Osnabrugge, G., Leedumrongwatthanakun, S. & Vellekoop, I. M. A convergent Born series for solving the inhomogeneous Helmholtz equation in arbitrarily large media. *Journal of computational physics* **322**, 113-124 (2016).
32 Yasuhiko, O. & Takeuchi, K. Bidirectional in-silico clearing approach for deep refractive-index tomography using a sparsely sampled transmission matrix. *Biomedical Optics Express* **15**, 5296-5313 (2024).
33 Yasuhiko, O. & Takeuchi, K. In-silico clearing approach for deep refractive index tomography by partial reconstruction and wave-backpropagation. *Light: Science & Applications* **12**, 101 (2023).
34 Yang, S., Kim, J., Swartz, M. E., Eberhart, J. K. & Chowdhury, S. DMD and microlens array as a switchable module for illumination angle scanning in optical diffraction tomography. *Biomedical Optics Express* **15**, 5932-5946 (2024).
35 Ivanov, V. Y., Sivokon, V. & Vorontsov, M. Phase retrieval from a set of intensity measurements: theory and experiment. *JOSA A* **9**, 1515-1524 (1992).
36 Grohs, P., Koppensteiner, S. & Rathmair, M. Phase retrieval: uniqueness and stability. *SIAM Review* **62**, 301-350 (2020).
37 Kumar, A. N., Short, K. W. & Piston, D. W. A motion correction framework for time series sequences in microscopy images. *Microscopy and Microanalysis* **19**, 433-450 (2013).
38 Mlodzianoski, M. J. *et al.* Sample drift correction in 3D fluorescence photoactivation localization microscopy. *Optics express* **19**, 15009-15019 (2011).

39. Bian, L. *et al.* Motion-corrected Fourier ptychography. *Biomedical optics express* **7**, 4543-4553 (2016).
40. Beckers, M. *et al.* Drift correction in ptychographic diffractive imaging. *Ultramicroscopy* **126**, 44-47 (2013).
41. Gao, Y. & Cao, L. Motion-resolved, reference-free holographic imaging via spatiotemporally regularized inversion. *Optica* **11**, 32-41 (2024).
42. Kellman, M., Chen, M., Phillips, Z. F., Lustig, M. & Waller, L. Motion-resolved quantitative phase imaging. *Biomedical optics express* **9**, 5456-5466 (2018).
43. Cao, R., Liu, F. L., Yeh, L.-H. & Waller, L. Dynamic structured illumination microscopy with a neural space-time model. *2022 IEEE International Conference on Computational Photography (ICCP)*, 1-12 (2022).
44. Saguy, A. *et al.* DBlink: Dynamic localization microscopy in super spatiotemporal resolution via deep learning. *Nature methods* **20**, 1939-1948 (2023).
45. Sankaran, J. *et al.* Simultaneous spatiotemporal super-resolution and multi-parametric fluorescence microscopy. *Nature communications* **12**, 1748 (2021).
46. Bohra, P., Pham, T.-a., Long, Y., Yoo, J. & Unser, M. Dynamic Fourier ptychography with deep spatiotemporal priors. *Inverse Problems* **39**, 064005 (2023).
47. Zhang, J. *et al.* Spatiotemporal coherent modulation imaging for dynamic quantitative phase and amplitude microscopy. *Optics Express* **29**, 38451-38464 (2021).
48. Priessner, M. *et al.* Content-aware frame interpolation (CAFI): Deep Learning-based temporal super-resolution for fast bioimaging. *Nature Methods* **21**, 322-330 (2024).
49. Tian, L. & Waller, L. Quantitative differential phase contrast imaging in an LED array microscope. *Optics express* **23**, 11394-11403 (2015).
50. Chen, M., Phillips, Z. F. & Waller, L. Quantitative differential phase contrast (DPC) microscopy with computational aberration correction. *Optics express* **26**, 32888-32899 (2018).
51. Sun, M., Wang, K., Mishra, Y. N., Qiu, S. & Heidrich, W. Space-time Fourier ptychography for in vivo quantitative phase imaging. *Optica* **11**, 1250-1260 (2024).
52. Meinhardt-Llopis, E. & Sánchez, J. Horn-schunck optical flow with a multi-scale strategy. *Image Processing on line* (2013).
53. Qi, M. & Heidrich, W. Scattering-aware Holographic PIV with Physics-based Motion Priors. *2023 IEEE International Conference on Computational Photography (ICCP), Madison, WI, USA, 2023*, 1-12 (2023).
54. Luo, H. *et al.* Dynamic multiplexed intensity diffraction tomography using a spatiotemporal regularization-driven disorder-invariant multilayer perceptron. *Optics Express* **32**, 39117-39133 (2024).
55. Ge, B. *et al.* Single-frame label-free cell tomography at speed of more than 10,000 volumes per second. *arXiv preprint arXiv:2202.03627* (2022).
56. Cao, R., Divekar, N. S., Nuñez, J. K., Upadhyayula, S. & Waller, L. Neural space–time model for dynamic multi-shot imaging. *Nature Methods*, 1-6 (2024).
57. Betzig, E. *et al.* Imaging intracellular fluorescent proteins at nanometer resolution. *science* **313**, 1642-1645 (2006).
58. Rust, M. J., Bates, M. & Zhuang, X. Sub-diffraction-limit imaging by stochastic optical reconstruction microscopy (STORM). *Nature methods* **3**, 793-796 (2006).

59  Gustafsson, M. G. Surpassing the lateral resolution limit by a factor of two using structured illumination microscopy. *Journal of microscopy* **198**, 82-87 (2000).
60  Gustafsson, M. G. Nonlinear structured-illumination microscopy: wide-field fluorescence imaging with theoretically unlimited resolution. *Proceedings of the National Academy of Sciences* **102**, 13081-13086 (2005).
61  Li, D. *et al.* Extended-resolution structured illumination imaging of endocytic and cytoskeletal dynamics. *Science* **349**, aab3500 (2015).
62  Chowdhury, S., Eldridge, W. J., Wax, A. & Izatt, J. A. Structured illumination microscopy for dual-modality 3D sub-diffraction resolution fluorescence and refractive-index reconstruction. *Biomedical optics express* **8**, 5776-5793 (2017).
63  Huang, X. *et al.* Fast, long-term, super-resolution imaging with Hessian structured illumination microscopy. *Nature biotechnology* **36**, 451-459 (2018).
64  Xu, X. *et al.* Ultra-high spatio-temporal resolution imaging with parallel acquisition-readout structured illumination microscopy (PAR-SIM). *Light: Science & Applications* **13**, 125 (2024).
65  Schaefer, L., Schuster, D. & Schaffer, e. J. Structured illumination microscopy: artefact analysis and reduction utilizing a parameter optimization approach. *Journal of microscopy* **216**, 165-174 (2004).
66  Förster, R., Wicker, K., Müller, W., Jost, A. & Heintzmann, R. Motion artefact detection in structured illumination microscopy for live cell imaging. *Optics Express* **24**, 22121-22134 (2016).
67  Xiong, B. *et al.* Recent developments in microfluidics for cell studies. *Advanced materials* **26**, 5525-5532 (2014).
68  Wu, J., Zheng, G. & Lee, L. M. Optical imaging techniques in microfluidics and their applications. *Lab on a Chip* **12**, 3566-3575 (2012).
69  Hung, P. J., Lee, P. J., Sabounchi, P., Lin, R. & Lee, L. P. Continuous perfusion microfluidic cell culture array for high-throughput cell-based assays. *Biotechnology and bioengineering* **89**, 1-8 (2005).
70  Kimura, H., Yamamoto, T., Sakai, H., Sakai, Y. & Fujii, T. An integrated microfluidic system for long-term perfusion culture and on-line monitoring of intestinal tissue models. *Lab on a Chip* **8**, 741-746 (2008).
71  Merola, F. *et al.* Tomographic flow cytometry by digital holography. *Light: Science & Applications* **6**, e16241-e16241 (2017).
72  Rees, P., Summers, H. D., Filby, A., Carpenter, A. E. & Doan, M. Imaging flow cytometry. *Nature Reviews Methods Primers* **2**, 86 (2022).
73  Guizar-Sicairos, M., Thurman, S. T. & Fienup, J. R. Efficient subpixel image registration algorithms. *Optics letters* **33**, 156-158 (2008).
74  Hill, D. L., Batchelor, P. G., Holden, M. & Hawkes, D. J. Medical image registration. *Physics in medicine & biology* **46**, R1 (2001).
75  Goodman, J. W. *Introduction to Fourier optics*.  (Roberts and Company publishers, 2005).
76  Saleh, B. E. & Teich, M. C. *Fundamentals of photonics*.  (john Wiley & sons, 2019).
77  Schmitt, U. & Louis, A. Efficient algorithms for the regularization of dynamic inverse problems: I. Theory. *Inverse Problems* **18**, 645 (2002).
78  Gradshteyn, I. S. & Ryzhik, I. M. *Table of integrals, series, and products*.  (Academic press, 2014).
79  Osher, S., Burger, M., Goldfarb, D., Xu, J. & Yin, W. An iterative regularization method for total variation-based image restoration. *Multiscale Modeling & Simulation* **4**, 460-489 (2005).

80  Perraudin, N., Kalofolias, V., Shuman, D. & Vandergheynst, P. UNLocBoX: A MATLAB convex optimization toolbox for proximal-splitting methods. *arXiv preprint arXiv:1402.0779* (2014).
81  Burger, W. & Burge, M. J. *Digital image processing: An algorithmic introduction*.  (Springer Nature, 2022).
82  Gross, H. Handbook of Optical Systems.  (2005).
83  Ziemczonok, M., Kuś, A., Wasylczyk, P. & Kujawińska, M. 3D-printed biological cell phantom for testing 3D quantitative phase imaging systems. *Scientific reports* **9**, 18872 (2019).
84  Krauze, W. *et al.* 3D scattering microphantom sample to assess quantitative accuracy in tomographic phase microscopy techniques. *Scientific Reports* **12**, 19586 (2022).